%% file: 0_main.tex
\documentclass[runningheads]{llncs}
\usepackage[T1]{fontenc}

\usepackage{etoolbox}
\usepackage[paperheight=235mm, paperwidth=155mm,textwidth=12.2cm,textheight=19.3cm,twoside=false]{geometry}

\usepackage{graphicx}
\usepackage{amsmath,amssymb}
\usepackage{hyperref}
\usepackage{cite}
\usepackage{csquotes}
\usepackage[capitalize]{cleveref}
\usepackage{xcolor}
\usepackage{microtype}
\usepackage{marvosym}
\usepackage{xspace}
\usepackage{subcaption}
\usepackage{booktabs}
\usepackage{rotating}
\usepackage{listings}
\usepackage{multicol}
\usepackage{dsfont}
\usepackage{etoolbox}
\usepackage{hyperref}

\usepackage{enumitem}
\setitemize{noitemsep,topsep=0pt,parsep=0pt,partopsep=0pt,leftmargin=12.0pt}
\setenumerate{noitemsep,topsep=0pt,parsep=0pt,partopsep=0pt}
\setdescription{noitemsep,topsep=0pt,parsep=0pt,partopsep=0pt,leftmargin=12.5pt}

\pretocmd{\subsubsection}{\vspace{-8pt}}{}{}

\makeatletter
\def\orcidID#1{\textsuperscript{\,\smash{\protect\raisebox{-1.25pt}{\href{http://orcid.org/#1}{\protect\includegraphics[scale=.8]{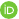}}}}}}
\makeatother

\usepackage{tikz}
\usetikzlibrary{shapes,positioning,arrows,arrows.meta,calc,automata,matrix,fit,backgrounds}
\tikzstyle{state}+=[minimum size = 6mm, inner sep=0,outer sep=1]
\colorlet{disabled}{lightgray}
\tikzstyle{state}=[draw,rectangle,inner sep=5pt,rounded corners=2pt]
\tikzstyle{action}=[font=\small,inner sep=0pt,outer sep=2pt]
\tikzstyle{actionnode}=[circle,draw=black,fill=black,minimum size=1mm,inner sep=0,outer sep=0]
\tikzstyle{actionedge}=[draw,-]
\tikzstyle{prob}=[font=\scriptsize,inner sep=0pt,outer sep=0pt,darkgray]
\tikzstyle{probedge}=[draw=darkgray,->]
\tikzstyle{directedge}=[draw,->]
\tikzset{chainarrow/.tip={Stealth[length=3pt]}}
\tikzset{>=chainarrow}
\usepackage{pgfplots}
\pgfplotsset{compat=1.18}

\input{0_macros}

\input{plots}

\let\llncssubparagraph\subparagraph
\let\subparagraph\paragraph
\usepackage{titlesec}
\let\subparagraph\llncssubparagraph

\titlespacing*{\subsubsection}{0pt}{1ex plus 0.5ex minus 0ex}{*1}
\titlespacing*{\paragraph}{0pt}{0.75ex plus 0.5ex minus 0ex}{*1}

\crefname{section}{Sec.}{Secs.}
\crefname{appendix}{App.}{Apps.}
\crefname{lemma}{Lem.}{Lemms.}
\crefname{theorem}{Thm.}{Thms.}
\crefname{corollary}{Cor.}{Cors.}
\crefname{equation}{Eq.}{Eqs.}
\crefname{figure}{Fig.}{Figs.}
\crefname{tabular}{Tab.}{Tabs.}
\crefname{algorithm}{Alg.}{Algs.}

\title{
UMB: A Unified Markov Binary Format\\ for Probabilistic Model Checking%
\thanks{The authors are ordered alphabetically. 
This work was initiated by \href{https://www.dagstuhl.de/24134}{Dagstuhl Meeting 24134} \enquote{Towards A Unified Interface For Modern Probabilistic Model Checking Tools}.
}\\(extended version)
}

\titlerunning{UMB: Unified Markov Binary Format for Probabilistic Model Checking}

\author{
Roman~Andriushchenko\inst{1}\orcidID{0000-0002-1286-934X}
\and Arnd~Hartmanns\inst{2}\orcidID{0000-0003-3268-8674}
\and Joshua~Jeppson\inst{3}\orcidID{0000-0002-8269-9489}
\and Sebastian~Junges\inst{4}\orcidID{0000-0003-0978-8466}
\and Tobias~Meggendorfer\inst{5}\orcidID{0000-0002-1712-2165}
\and David~Parker\inst{6}\orcidID{0000-0003-4137-8862}
\and Tim~Quatmann\inst{7}\orcidID{0000-0002-2843-5511}
\and Maximilian~Weininger\inst{8}\orcidID{0000-0002-0163-2152}
}
\authorrunning{R.\ Andriushchenko et al.} %
\institute{
Brno University of Technology, Brno, Czechia
\and University of Twente, Enschede, The Netherlands
\and Utah State University, Logan, Utah, USA
\and Radboud University, Nijmegen, The Netherlands
\and Lancaster University Leipzig, Leipzig, Germany
\and University of Oxford, Oxford, UK
\and RWTH Aachen University, Aachen, Germany
\and Ruhr-University Bochum, Bochum, Germany
\vspace{-3.5pt}
}

\hypersetup{
  pdftitle={UMB: A Unified Markov Binary Format for Probabilistic Model Checking},
  pdfauthor={Roman Andriushchenko, Arnd Hartmanns, Joshua Jeppson, Sebastian Junges, Tobias Meggendorfer, David Parker, Tim Quatmann, Maximilian Weininger}
}

\begin{document}

\maketitle

\begin{abstract}
This paper presents the unified Markov binary (UMB) format, an efficient, extensible, and well-supported explicit-state file format for representing a wide range of probabilistic systems.
UMB addresses the problem that, while probabilistic model checking tools
often support common high-level modelling languages,
there is no effective mechanism for exchanging low-level model representations.
In practice, textual, tool-specific formats are used,
hampering interoperability and resulting in large overheads in writing and reading model files.
UMB provides a clean, unified, and efficient solution, based on a general underlying mathematical model,
and encoded using a small set of bit-level primitive data structures.
The format has already been adopted by prominent tools
and comes with a convenient Python library for reading, manipulating, creating,
and validating models, plus infrastructure for cross-tool installation and
continuous validation. We report on both the efficiency of the file format
and the new practical use cases that it facilitates.
    
\end{abstract}

\input{1_intro}

\input{2_math_repr}

\input{3_format}

\input{4_tool_support}
\input{5_use_cases_experiments}

\input{9_conclusion}

\bibliography{references.bib}
\bibliographystyle{splncs04}
\clearpage
\appendix
\crefalias{section}{appendix} 
\crefalias{subsection}{appendix} 
\crefalias{subsubsection}{appendix}

\appendix
\section*{Appendix}
\input{10_appx_umbt}
\input{10_appx_models}

\input{10_appx_experiments}

\end{document}

%% file: 0_macros.tex
\newcommand{\ats}{\mathcal{S}} %
\renewcommand{\ts}{\mathcal{T}} %
\newcommand{\anns}{\mathcal{A}} %

\newcommand{\states}{\mathsf{S}} %
\newcommand{\choices}{\mathsf{C}} %
\newcommand{\branches}{\mathsf{B}} %
\newcommand{\entities}{\mathsf{E}} %

\newcommand{\ap}{\mathsf{AP}} %

\newcommand{\one}{\mathds{1}}

\usepackage{xargs}
\PassOptionsToPackage{disable}{todonotes}
\usepackage[textwidth=25mm,
	backgroundcolor=white,
    tickmarkheight=2pt,
	textsize=tiny]{todonotes}
\usepackage{silence}
\newcommandx{\todomacro}[4][4=]{%
    \todo[bordercolor=#2, linecolor=#2, , backgroundcolor=#2!4, #4]{\textcolor{#2}{\textbf{#1:} #3}}%
}
\newcommandx{\dave}[2][2=]{\todomacro{DP}{violet}{#1}[#2]} %
\newcommandx{\mw}[2][2=]{\todomacro{MW}{blue}{#1}[#2]}
\newcommandx{\sj}[2][2=]{\todomacro{SJ}{cyan}{#1}[#2]}
\newcommandx{\tm}[2][2=]{\todomacro{TM}{orange}{#1}[#2]}
\newcommandx{\jj}[2][2=]{\todomacro{JJ}{black}{#1}[#2]}
\newcommandx{\ra}[2][2=]{\todomacro{RA}{green}{#1}[#2]}
\newcommandx{\ah}[2][2=]{\todomacro{AH}{yellow!70!black}{#1}[#2]}
\newcommandx{\tq}[2][2=]{\todomacro{TQ}{red}{#1}[#2]}

\newcommand{\tool}[1]{\textsc{#1}\xspace}
\newcommand{\lang}[1]{\textsc{#1}\xspace}
\newcommand*{\jani}{\lang{Jani}}

\newcommand*{\UMB}{\lang{UMB}}

\newcommand*{\mcsta}{\tool{mcsta}}

\newcommand*{\toolset}{\tool{Modest Toolset}}

\newcommand*{\paynt}{\tool{Paynt}}
\newcommand*{\pet}{\tool{Pet}}
\newcommand*{\prism}{\tool{Prism}}

\newcommand*{\stamina}{\tool{Stamina}}
\newcommand*{\storm}{\tool{Storm}}

\newcommand*{\stormvogel}{\tool{Stormvogel}}

\newcommand*{\umbi}{\tool{umbi}}
\newcommand*{\UMBI}{\tool{umbi}}

%% file: plots.tex
\newtoggle{showplots}
\toggletrue{showplots}

\definecolor{plotred}{RGB}{255,0,0}
\definecolor{plotgreen}{RGB}{0,255,0}
\colorlet{plotblack}{black}
\colorlet{plotblue}{blue}
\colorlet{plotlightblue}{plotblue!60!white}
\definecolor{plotyellow}{RGB}{230,230,0}
\colorlet{plotcyan}{cyan!70!white}
\definecolor{plotorange}{RGB}{255,127,0}
\definecolor{plotpink}{RGB}{255,0,255}
\colorlet{plotgray}{gray!70!white}
\definecolor{plotdarkgray}{RGB}{128,128,128}
\definecolor{plotdarkred}{RGB}{128,0,0}
\definecolor{plotgreenyellow}{RGB}{128,128,0}
\definecolor{plotdarkgreen}{RGB}{0,128,0}
\definecolor{plotpurple}{RGB}{128,0,128}
\definecolor{plotteal}{RGB}{0,128,128}
\definecolor{plotdarkblue}{RGB}{0,0,128}
\definecolor{plotlightred}{RGB}{205,92,92}

\newcommand{\numinstances}{36}
\newcommand{\quantileplotxlabel}{number of models (out of \numinstances)}
\newcommand{\quantileplotylabel}{runtime (seconds)}
\newlength{\quantileplotwidth}
\newlength{\quantileplotheight}
\newcommand{\quantileplotlegendcols}{1}
\newcommand{\standardquantileplotlinestyle}{ultra thick}
\newcommand{\quantileplotlinestyle}{\standardquantileplotlinestyle}

\newcommand{\quantileplot}[8]{%
    \resizebox{\linewidth}{!}{%
    \begin{tikzpicture}
    \begin{axis}[
        width=\quantileplotwidth,
        height=\quantileplotheight,
        xmin=#4,
        xmax=#5,
        ymin=#6,
        ymax=#7,
        ymajorgrids,
        ymode=log,
        axis x line=bottom,
        axis y line=left,
        unbounded coords=discard,filter discard warning=false, %
        xlabel=\quantileplotxlabel,
        xlabel style={font=\scriptsize,yshift=4pt},%
        ylabel=\quantileplotylabel,
        ylabel style={font=\scriptsize,yshift=-7pt},%
        yticklabel style={font=\scriptsize},
        scaled y ticks=false,
        xticklabel style={font=\scriptsize},
        legend columns=\quantileplotlegendcols,
        legend pos={#8},
        legend style={font=\scriptsize,nodes={scale=0.75, transform shape},inner sep=1pt,row sep=0pt,draw=none},
        /pgfplots/legend image code/.code={\draw[mark repeat=2,mark phase=2,##1] plot coordinates {(0cm,0cm) (0.3cm,0cm)};}, %
        every axis plot/.append style={\quantileplotlinestyle},
        legend cell align={left}
    ]
    \iftoggle{showplots}{
        \foreach \tool\color in {#2}{%
            \edef\loopbody{
                \noexpand\addplot[\color] table [x=n,y=\tool, col sep=tab] {#1};
            }
            \loopbody
        }
        \legend{#3}}{\node[anchor=south west, align=center, red] {\huge NOT\\ COMPILED};}
    \end{axis}
    \end{tikzpicture}}%
}

\newcommand{\quantileplotsize}[8]{{%
\renewcommand{\quantileplotylabel}{size (bytes)}%
\quantileplot{#1}{#2}{#3}{#4}{#5}{#6}{#7}{#8}%
}}

\newlength\scatterplotsize
\newcommand{\scatterplot}[6]{%
    \begin{tikzpicture}
    \begin{axis}[
    width=\scatterplotsize,
    height=\scatterplotsize,
    axis equal image,
    xmin=0.05,
    ymin=0.05,
    ymax=7200,
    xmax=7200,
    xmode=log,
    ymode=log,
    axis x line=bottom,
    axis y line=left,
     xtick={1,10,100,1000},
     ytick={1,10,100,1000},
    xlabel={#3},
    xlabel style={font=\scriptsize},%
    ylabel={#5},
    ylabel style={font=\scriptsize,yshift=-5pt},%
    yticklabel style={font=\scriptsize},
    xticklabel style={font=\scriptsize},%
    set layers,
    mark layer=axis background
    ]

    \ifthenelse{\NOT\equal{#6}{false}}{\addlegendimage{empty legend}}{}
    \iftoggle{showplots}{\addplot[
        only marks, mark=x,plotblue
    ]%
    table [col sep=tab,x=#2,y=#4] {#1};
    }{\node[anchor=south west, align=center, red] {\huge NOT\\ COMPILED};}
    \ifthenelse{\NOT\equal{#6}{false}}{%
    }{}
    \addplot[no marks] coordinates {(0.01,0.01) (7200,7200) };
    \addplot[no marks, densely dotted] coordinates {(0.01,0.02) (3600,7200)};
    \addplot[no marks, densely dotted] coordinates {(0.02,0.01) (7200,3600)};
    \end{axis}
    \end{tikzpicture}
}

%% file: 1_intro.tex
\section{Introduction}

Probabilistic model checking (PMC) is a technique for formal modelling and analysis of stochastic systems~\cite{BK08,BAFK18,KNP25}.
Several mature PMC tools~\cite{KNP11,HH14,HJKQV22,ACJKS21,JVI+23,MW24} have been developed over the past decades.
They support models ranging from basic Markov chains and Markov decision processes (MDPs)~\cite{HHH+19,BHK+20,HJQW26} to extensions such as partially-observable MDPs and stochastic games~\cite{ABB+24}. 
Since the features and strengths of these tools differ,
and because they are commonly incorporated into toolchains,
efficient and reliable interconnectivity is crucial.

As a \emph{high-level} modelling formalism,
the textual PRISM language~\cite{prismlang} has become a de facto standard for probabilistic modelling and model checking, supported by various tools and covering many types of models.
The JANI~\cite{BDH+17} JSON-based format facilitates the interchange of high-level probabilistic model descriptions between tools.
Internally, \emph{low-level} storage and solution of models in probabilistic model checkers is now very often \emph{explicit-state}, i.e., states, transitions and their probabilities are explicitly enumerated in memory.
There is a distinct lack of effective \emph{interfaces} for such model representations between tools.
In practice, it is usual to either (i)~re-encode the models into high-level formats such as the PRISM language, creating a large overhead in storage and parsing, or (ii)~use simple, tool-specific textual formats~\cite{tra,drn},
which tend to be ad hoc, limited to certain model types,
and also inefficient to generate and parse.

To address this need for a universal low-level model exchange format, we present the \emph{unified Markov binary} (\UMB) format, an efficient, extensible and well-supported file format for the explicit-state representation of a wide range of probabilistic systems.
Working with probabilistic model checking tool developers and users,
we have devised a general-purpose model encoding, along with a clearly defined mapping to the format
from the majority of (discrete space, Markovian) model types in common use.
This includes Markov chains, Markov decision processes,
and their extensions with with continuous time, partial observability,
uncertain (interval) transition probabilities, and multiple players.

\subsubsection{The \UMB format.}
We formalise our model encoding with a unifying mathematical representation (see \cref{sec:math-repr})
in which transitions between \emph{states} are layered into \emph{choices} and \emph{branches}.
These three fundamental objects are then augmented with \emph{annotations}, capturing further model information, such as transition probabilities, atomic propositions, or rewards.

Building on this formalism, we define the \UMB file format (see \cref{sec:format}),
which combines structured metadata and binary files to encode the model data.
For the latter, we focus on a small set of bit-level defined primitive data structures,
aiming to strike a balance between simplicity and efficiency/compactness.
We align with established storage schemes (such as the compressed sparse row matrix format) to reduce data transformation overheads. %
Our experiments in \cref{sec:empirical} illustrate the resulting gains in processing times. %

Moreover, the format has been designed with extensibility in mind.
For example, it is straightforward to use combinations of modelling features that still lack tool support, e.g., partially observable stochastic games,
or interval continuous-time Markov chains.
Further, while we \emph{currently} choose to exclude models that include
\emph{symbolic} expressions (e.g., probabilistic timed automata, which comprise guard expressions over real-valued clock variables~\cite{KNSS02}), 
the file format is flexible enough to accommodate these via, e.g., JSON-based annotations. %

\subsubsection{The \UMB ecosystem.}
\UMB is supported by the most prominent probabilistic model checking tools, including \prism~\cite{KNP11}, \mcsta~\cite{HH15} of the \toolset~\cite{HH14}, and \storm~\cite{HJKQV22}, as well as tools reusing their parsers, e.g., \paynt~\cite{ACJKS21} and \stamina~\cite{JVI+23}. %
Furthermore, it is accompanied by the \umbi Python library, which acts as a reference implementation, offers methods for creating, modifying and analysing models, and serves as an interface to \UMB, e.g., for the visualisation library  \stormvogel~\cite{LMHVJ25}.
Finally, the \emph{\UMB observatory} 
ensures alignment of the different tools supporting \UMB by providing a shared test environment and continuous integration.
We describe UMB tooling in more detail in \cref{sec:ecosystem}.
Then, in \cref{sec:tooldemo}, we discuss some of the benefits that \UMB support provides,
highlighting a number of novel \emph{use cases} for the format,
including tool chains that combine the strengths of different tools and rapid prototyping.

\subsubsection{Related file formats.}
Established modelling formalisms like the PRISM language and JANI
serve a different purpose, targetting \emph{high-level} model descriptions
at the level of components and variables, defining model behaviour using symbolic expressions.
As such, they are poorly suited to encode explicit-state, \emph{low-level} models which do not lend themselves to concise, symbolic representation. %

Existing explicit-state file formats for probabilistic model checking
tend to be tool-specific, text-based and with various limitations.
PRISM uses the \texttt{.tra} format~\cite{tra} to write/read the core transition behaviour.
Other model entities, such as rewards, need to be
stored in separate files, and the files lack metadata, instead relying on
automatic inference of model type or numerical precision.
\storm offers the \texttt{.drn} format~\cite{drn},
which provides a more general, structured textual format, but as a result is more verbose than \texttt{.tra}.
Both remain somewhat ad-hoc formats and are very inefficient for reading and writing large models.

We also note related file formats from other areas.
BCG is a proprietary format of the CADP toolset~\cite{GLMS13},
primarily for labelled transition systems, but with the facility to add probability labels,
albeit in a much more limited fashion than is possible with \UMB.
BCG uses a very compact binary encoding.
The MoXI~\cite{JND+24} model exchange format
focuses on symbolic encodings of non-probabilistic models.

%% file: 2_math_repr.tex
\section{Unified Mathematical Representation}\label{sec:math-repr}

Underlying our file format is the unified mathematical representation of \emph{annotated transition systems} (ATSs).
In an ATS, transitions between \emph{states} comprise two layers, \emph{choices} and \emph{branches};
a wide range of \emph{annotations} can be applied to all of these.
Many commonly employed models,
including Markov decision processes, Markov chains,
and a wide range of others supported by probabilistic model checkers~\cite{ABB+24}
can be captured naturally in ATSs.
Furthermore, the formalism is designed to map cleanly to a generalisable implementation as a file format.

\begin{definition}[Annotated Transition System]\label{def:ats}
	A (discrete-space, time-homogenous) \emph{annotated transition system} (ATS) is a pair $\ats = (\ts, \anns)$ of a \emph{transition system} $\ts$ and a set of \emph{annotations} $\anns$.
    A transition system is a triple $\ts = (\states,\choices,\branches)$, consisting of finite sets of \emph{states}~$\states$, \emph{choices}~$\choices$, and \emph{branches}~$\branches$.
    Each state $s\in \states$ has a (possibly empty) set of choices $\choices_s \subseteq \choices$, and each choice $c \in \choices$ has a (possibly empty) set of branches $\branches_c \subseteq \branches$.
    These sets are disjoint, i.e.\ $\choices = \biguplus_{s\in\states} \choices_s$ and $\branches = \biguplus_{c\in\choices} \branches_c$.
    Further, every branch $b\in\branches$ defines a (not necessarily unique) successor state $s_b\in\states$.
	An \emph{annotation} is a mapping $\entities \to X$ from an \emph{entity} $\entities \in \{\states,\choices,\branches\}$ of the transition system to some domain $X$.
\end{definition}
\begin{figure}[t]
	\centering
    \begin{subfigure}[b]{0.45\textwidth}
        \centering
	\begin{tikzpicture}[auto,xscale=1]
    	\node[state] at (0,0) (s0) {$s_0$};
    	\node[state] at (-2,-2) (s1) {$s_1$};
    	\node[state] at (0,-2) (s2) {$s_2$};
    	\node[state] at (2,-2) (s3) {$s_3$};
    
        \node[actionnode] at (-1,-1) (c0) {};
    	\node[actionnode] at (1,-1) (c1) {};
    	\node[actionnode] at (-2.5,-1) (c2) {};
    	\node[actionnode] at (-0.75,-2.25) (c3) {};
    	\node[actionnode] at (1.25,-2.25) (c4) {};
    
    	\path[actionedge]
    		(s0) edge[swap] node[action] {$c_0$} (c0)
    		(s0) edge node[action] {$c_1$} (c1)
    		(s1) edge[swap,bend right] node[action,pos=0.67] {$c_2$} (c2)
    		(s2) edge[swap,out=160,in=90] node[action] {$c_3$} (c3)
    		(s3) edge[swap,out=160,in=90] node[action] {$c_4$} (c4)
    	;
    	\path[probedge]
    		(c0) edge[swap] node[prob] {$b_0$} (s1)
    		(c0) edge node[prob] {$b_1$} (s2)
    
    		(c1) edge[swap] node[prob] {$b_2$} (s2)
    		(c1) edge node[prob] {$b_3$} (s3)
    
    		(c2) edge[swap,bend right] node[prob] {$b_4$} (s1)
    		(c2) edge[bend left] node[prob] {$b_5$} (s0)
    
    		(c3) edge[swap,out=-70,in=-120] node[prob,yshift=-1pt,overlay] {$b_6$} (s2)
    		(c4) edge[swap,out=-70,in=-120] node[prob,yshift=-1pt,overlay] {$b_7$} (s3)
    	;
	\end{tikzpicture}
        \caption{}
        \label{fig:example-ts}
    \end{subfigure}%
    \begin{subfigure}[b]{0.45\textwidth}
        \centering
        \begin{tikzpicture}[auto,xscale=1]
    	\node[state] at (0,0) (s0) {$s_0$};
    	\node[state] at (-2,-2) (s1) {$s_1$};
    	\node[state] at (0,-2) (s2) {$s_2$};
    	\node[state] at (2,-2) (s3) {$s_3$};
    
        \draw[<-] (s0.east) -- ++(0.4,0);
        
    	\node[actionnode] at (-1,-1) (c0) {};
    	\node[actionnode] at (1,-1) (c1) {};
    	\node[actionnode] at (-2.5,-1) (c2) {};
    	\node[actionnode] at (-0.75,-2.25) (c3) {};
    	\node[actionnode] at (1.25,-2.25) (c4) {};
        
    	\path[actionedge]
    		(s0) edge[swap] node[action] {$a$} (c0)
    		(s0) edge node[action] {$b$} (c1)
    		(s1) edge[swap,bend right] node[action,pos=0.67] {$a$} (c2)
    		(s2) edge[swap,out=160,in=90] node[action] {$a$} (c3)
    		(s3) edge[swap,out=160,in=90] node[action] {$a$} (c4)
    	;
    	\path[probedge]
    		(c0) edge[swap] node[prob] {$0.5$} (s1)
    		(c0) edge node[prob] {$0.5$} (s2)
    
    		(c1) edge[swap] node[prob] {$0.2$} (s2)
    		(c1) edge node[prob] {$0.8$} (s3)
    
    		(c2) edge[swap,bend right] node[prob] {$0.1$} (s1)
    		(c2) edge[bend left] node[prob] {$0.9$} (s0)
    
    		(c3) edge[swap,out=-70,in=-120] node[prob,yshift=-2pt,overlay] {$1$} (s2)
    		(c4) edge[swap,out=-70,in=-120] node[prob,yshift=-2pt,overlay] {$1$} (s3)
    	;
	\end{tikzpicture}
        \caption{}
        \label{fig:example-mdp}
    \end{subfigure}
    \vspace*{-0.7em}
	\caption{
		An annotated transition system (left) and an MDP resulting from annotating the transition system with branch probabilities and (choice) actions (right).
	} \label{fig:prelim-example}
\end{figure}
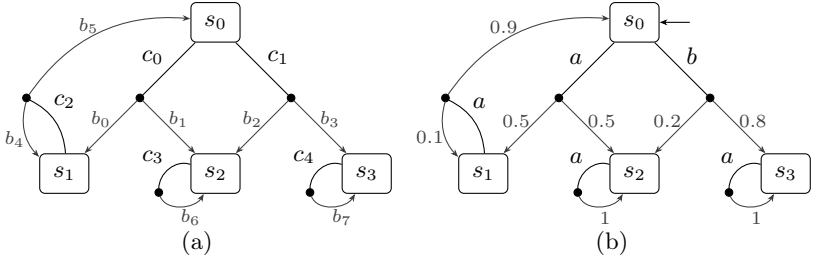
\Cref{fig:example-ts} shows a transition system
with 4 states, 5 choices and 8 branches.
We have $\choices_{s_0} = \{c_0,c_1\}$, $\branches_{c_1} = \{b_2,b_3\}$, and $s_{b_3} = s_3$.
The co-domain $X$ of annotations can be integers~$\mathbb{Z}$, reals $\mathbb{R}$, probabilities $[0,1]\subset \mathbb{R}$, Booleans $\{0,1\}$ or any other arbitrary set needed to define the model.
Annotations capture both core model data
(e.g., transition probabilities or state to player mappings in games) and other labellings needed for model checking (e.g., atomic propositions or rewards).

\subsubsection{Representing MDPs.}
An ATS can capture finite systems with Markovian dynamics, i.e., in which the definition of possible transitions to the next state depends only on the current state. It also naturally supports a combination of both nondeterministic and probabilistic choices.
As a first illustration of this, we describe an instantiation of ATSs
to Markov decision processes (MDPs)~\cite{Put94},
a widely used model in probabilistic model checking and other fields such as artificial intelligence and control theory.
MDPs are the simplest probabilistic model where all components of Definition~\ref{def:ats} are non-trivial.

\begin{definition}[Markov decision process]\label{def:mdp-classic}
    An MDP is a tuple $(\states, s_0, A, P)$ where
    $\states$ is a set of states,
    $s_0\in\states$ is an initial state,
    $A$ is a set of actions,
    and $P\colon \states\times A \to [0,1]$ is a transition probability function.
\end{definition}
We can naturally encode an MDP as a transition system $\ts$ and annotation set~$\anns$, where the latter includes mappings of the form
$\states \to \{0,1\}$, $\choices \to A$, and $\branches \to [0,1]$,
representing initial states, actions, and transition probabilities, respectively.
To match \Cref{def:mdp-classic}, we can require that a single
state is marked as initial and choice actions are distinct within a state,
but more general representations could also be permitted. %
\Cref{fig:example-mdp} shows how the example transition system above
is annotated to represent a 4-state MDP with two actions $A=\{a,b\}$.

\subsubsection{Property-related annotations.}
MDPs are typically used in conjunction with \emph{rewards} (or, conversely, costs),
most frequently assigned to state-action pairs ($\states\times A \to \mathbb{R}$),
but also to state-action-state triples ($\states\times A\times\states \to \mathbb{R}$) or states ($\states \to \mathbb{R}$).
All three are easily encoded as annotations for
choices, branches and states of an ATS, respectively.
MDPs are also often augmented with a labelling function that maps states to a set of \emph{atomic propositions} $\ap$,
allowing the specification of temporal logic properties.
We capture these with a set of Boolean-valued annotations
$\{f_a\colon\states \to \{0,1\} \mid a\in\ap\}$.

Information from a high-level model description can also be attached.
A good example is mapping states to \emph{valuations}, i.e., the values of \emph{variables}/\emph{factors}
(e.g., position, velocity, or protocol state) that make up each state.
These can be used in property specification, for debugging/validating model checking results,
and for generating concise, understandable policies~\cite{DJLMT15,AJK+21}.

\subsubsection*{Other model classes.}
Various other common classes of models can be encoded as ATSs, too,
illustrating the generality of our approach:
\begin{itemize}
\item 
\emph{Discrete-time Markov chains} (DTMCs) can be represented as MDPs with one choice per state.
For \emph{continuous-time Markov chains} (CTMCs), we adopt the common approach of treating them as a DTMC (the \enquote{embedded Markov chain})
with an \emph{exit rate} for each state, stored as an annotation $\states\to\mathbb{R}_{\geq 0}$.
For both types of Markov chains, actions become branch annotations.

\item 
\emph{Markov automata} (MA)~\cite{EHZ10} generalise CTMCs with nondeterministic choices.
They are naturally stored as a combination of the CTMC and MDP encodings, plus a Boolean annotation indicating which states are Markovian. %

\item 
\emph{Partially observable} models, such as POMDPs~\cite{KLC98}, add \emph{observations} $O$ made by an agent as the model evolves.
These are typically assigned to action-successor tuples $Z\colon A\times\states\to O$ or to states $Z\colon\states\to O$, which is representable as annotations of branches and states, respectively.
\emph{Stochastic} observations are also considered, where $Z$ instead maps to distributions over observations.

\item 
\emph{Interval models} are a common way to capture epistemic uncertainty in probabilistic systems. Interval Markov chains, interval MDPs~\cite{GLD00}, and even interval POMDPs, are directly supported in ATSs
by changing the type of the branch probability annotation from $[0,1]$ to $[0,1]^2$.

\item 
\emph{Stochastic games}~\cite{Sha53} generalise MDPs by allowing multiple \emph{players} to resolve
how choices are taken in each state.
For turn-based games~\cite{Con92}, we add an annotation mapping states to players, indicating control.
For concurrent games~\cite{Sha53}, an annotation of choices with tuples of players would be used.

\item 
Lastly, although our focus is Markovian models,
\emph{non-probabilistic} models are easily supported, too;
e.g. a labelled transition system (LTS,~\cite{BK08}) is a special MDP
with only one branch per choice and no branch probability annotation.
\end{itemize}
Note that the modelling features above are all orthogonal, so we can easily represent
\enquote{exotic} combinations, e.g., partially observable stochastic games or interval Markov automata,
greatly simplifying the process of adding new model variants.
This flexibility and extendibility are key benefits of our formalism.

%% file: 3_format.tex
\section{The \UMB File Format}\label{sec:format}
The \UMB format is the product of detailed discussions among probabilistic model checking tool developers and users, aiming for a \emph{unified}, widely supported approach to cleanly represent a broad range of probabilistic models. %
We now provide an overview of its practical details, with a simple, illustrative example in \cref{app:example}.
The complete technical specification of \UMB{} is available online at\\[2pt]
\centerline{\url{https://github.com/pmc-tools/umb/blob/main/spec.md}.}

A \texttt{.umb} file is tar file (optionally compressed)
containing a set of files with well-defined names in well-defined locations. 
A single JSON index file \texttt{index.json} defines metadata
describing, in particular, the type of the model, basic statistics, and what annotations are included.
This also acts as an index to the remaining contents of the tar file, which all are binary files.
These are of only three different kinds:
\texttt{TO1} files, representing integer-indexed arrays of fixed-size values;
\texttt{SEQ} files, containing lists of potentially variable-sized values;
and \texttt{CSR} files, which act as indices into other files,
using the standard \emph{compressed sparse row} storage scheme~\cite{Saa03}.
Together, this provides constant-time look-up of, for example, the branches of an ATS indexed by state/choice
(matching typically-used storage schemes)
and annotations with values of variable sizes such as strings.

The core \emph{entities} of a UMB-encoded model %
are
\emph{states}, \emph{choices} and \emph{branches}.
In the file format, we also treat \emph{observations} and \emph{players} as first-class model entities.
We assume that all five entity types have a fixed, integer indexing
starting from zero. %
\UMB also defines a small set of \emph{types}, including Booleans,
numeric types (integers, doubles, and rationals, of configurable sizes), intervals and strings.

\subsubsection{Core model data.}
A first, mostly compulsory, set of binary files define the basic structure of the transition system.
The underlying graph structure of the model
is captured (using \texttt{CSR} storage) by the relation between states, choices, and branches.
Another core annotation, used in almost all models,
is a \texttt{TO1} mapping from branches to probability values of a suitable numeric type.
As discussed in \cref{sec:math-repr},
the remaining core model information for a particular model class
then comprises annotations of model entities by values of a particular type,
representing, for example, actions, exit rates, observations, or players.

\subsubsection{Annotations.}
Beyond the core model data, \UMB provides a mechanism for attaching
arbitrary \emph{annotations} of any model entity
with values from any supported type.
These are given a unique \emph{ID}, for internal use,
and an \emph{alias}, for linking to external property descriptions,
and are organised into named \emph{groups}.
Two pre-defined groups exist for
\emph{rewards} (of numeric types, attached to states/choices/branches)
and \emph{atomic propositions} (Boolean-valued, attached to states);
others can be freely added.
Support is also provided for annotations of \emph{distributions} over basic types, a mechanism also used for stochastic observations in POMDPs.

\subsubsection{Valuations.}
A model can also optionally include, for each entity type,
a set of \emph{valuations}: maps from the entities
to tuples of values for a set of \emph{variables}.
The primary usage is to provide a definition of the meaning of states
(and, for partially observable models, observations),
with respect to a set of variables (or state factors) in a high-level model description.
For compactness, each valuation is bit-packed,
following a configurable storage scheme specified in the metadata.

\subsubsection{Design goals.}
Design decisions for the format have been driven by several overarching goals.
One is to support not only the full range of model classes currently in common use,
but also a wide range of \emph{extensions} and \emph{generalisations}.
The annotation-based approach to model storage enables this,
as does the \UMB approach to storing model metadata:
for example, rather than encoding the specific model type (e.g., MDP, DTMC),
we specify each modelling feature (e.g., nondeterminism, observations, players) separately (see \cref{app:example} for an example).
The flexibility of the format also facilitates adding new modelling features.

Secondly, \UMB aims for a sweet spot between \emph{usability} and \emph{efficiency},
where usability applies to both tool developers and tool users,
and efficiency covers both export/import speed and compactness of storage.
For example, the storage schemes for model data, using CSRs and arrays of primitive types,
allow tools to work directly on memory-mapped UMB files; but we do not optimise the bit encoding of every value type,
to allow for easier import/export.
Similarly, we clearly define default values for absent binary files,
meaning that the structure of the underlying ATS formalism does not result in storage of redundant information (for example, a Markov chain's trivial choice definitions can be omitted).

Finally, we note that models with \emph{symbolic} expressions
(such as clock guards in real-time models,
or rational polynomial probabilities in parametric MDPs) are not yet supported by \UMB.
While mathematically simple at the ATS level,
we postpone finalising an efficient, future-proof storage scheme to the next version,
for example with annotations in JSON, which the format already supports.

%% file: 4_tool_support.tex
\section{The \UMB Ecosystem}\label{sec:ecosystem}

\subsubsection{\UMBI Python Library.} %
\umbi, available at \href{https://github.com/pmc-tools/umbi}{github.com/pmc-tools/umbi}, is a lightweight Python library and command-line interface for reading, manipulating, creating, and validating UMB files.
It makes interacting with the model straightforward by providing low-level file access via concise, well-documented primitives.
The library features a modular and extensible design with small, focused modules and clear extension points for custom adapters or future format extensions.

Beyond file-level operations, \umbi offers a format-agnostic abstraction layer that enables users to manipulate an ATS without understanding UMB internals. The ATS is represented as a simple object model, with entities and their annotations accessible via iterable collections and query helpers, following a first-state-next-state API. By hiding file layout details, this abstraction ensures that code written against the API works regardless of the underlying storage format and remains valid in the face of format changes, streamlining tool integration and allowing a focus on models and algorithms rather than file format specifics.

\subsubsection{Tool support.} %
\UMB is supported by the majority of current PMC tools.

\prism, as of version 4.10, provides \UMB import and export for all models supported by both the tool and the format, notably Markov chains, MDPs and POMDPs, interval variants of all these, and LTSs.
This includes rewards, atomic propositions, valuations for states/observations,
and either floating point or exact arithmetic models.
\tool{PRISM-games}, as of version 3.2.2, provides the same functionality for turn-based stochastic games.
Support for reading and writing \UMB files has been implemented as a standalone \texttt{umbj} Java library that is available at \href{https://github.com/pmc-tools/umbj}{github.com/pmc-tools/umbj}.

\storm (version 1.12) integrates \UMB support (either compressed or uncompressed) for import and export of Markov automata, stochastic games, POMDPs, and all their special cases, including interval- or rational-valued transitions and including rich annotations on states and observations. The support is deeply embedded and thus \storm{} provides an interface also to, e.g., fault trees or Petri nets, or can be used to provide, e.g., bisimulation quotients of UMB models.

In the \toolset, the \mcsta probabilistic model checker has been rebuilt around the UMB format.
Its internal data structures match the files and binary representation of UMB, allowing \mcsta to ``load'' UMB files by simply mapping them into memory via \texttt{mmap}.
It supports reading \UMB files encoding LTSs, DTMCs, CTMCs, MDPs, and MAs, and can export such models
given in \jani or the \lang{Modest} language~\cite{BDHK06,HHHK13} to UMB.
As working with unchecked pointers/array indexing on untrusted \UMB data could easily create security vulnerabilities, \mcsta includes a strict \UMB validator that thoroughly checks the metadata and binary files for semantic consistency and data ranges. %

The tools~\paynt, \stormvogel, and \stamina use parts of \storm, and thus can also now import and export \texttt{.umb} files.
Similarly, \pet, which uses \prism's input parser, will support \UMB{} through \prism's \texttt{umbj} library.

\subsubsection{\UMB observatory.} %
To ensure alignment of \UMBI and the different tools supporting \UMB, we provide a shared test environment (the \emph{\UMB observatory} at \href{https://github.com/pmc-tools/umb-observatory}{github.com/pmc-tools/umb-observatory}) and continuous integration.
The observatory is realised as a container with the latest versions of \UMB-compatible tools and \UMBI installed. 
We provide thin Python wrappers around the command-line tools to simplify using and configuring them. 
The observatory includes tests for tools, tool-\UMBI compatibility, and cross-tool compatibility. 
They currently test the exception-free write-read-write-read paths through the tools along with some semantic checks.
The tests already demonstrated their benefits during development by detecting many subtle differences between parsers and exporters. Finally, a local Jupyter server with a simple notebook simplifies access to rerun individual tests.
The container is built every night to include the latest tool versions and runs the tests.

%% file: 5_use_cases_experiments.tex
\section{Applications for \UMB}
\label{sec:tooldemo}

\subsubsection{Explicit-state model communication.}
Interfacing with model checkers via high-level description languages like \prism or \jani is widespread in tools and benchmarks, even when explicitly listing states rather than providing high-level descriptions.
Examples include random benchmark generation~\cite{AEKSW22}, 
converters from ASCII to \jani~\cite{GHHKS20}, \lang{Julia} to \lang{PRISM} for POMDPs~\cite{HFRSL24}, 
\href{https://www.prismmodelchecker.org/bib.php?sort=subject\#connections}{and many others}.
We conjecture that this practice is partially due to the prominence of high-level description languages, their relative ease of use and their tool-independent support, while explicit-state formats like \texttt{.tra} and \texttt{.drn} remain less well known. %
However, encoding a transition system explicitly in high-level languages is highly ineffective: It massively blows up the representation, can obscure any structure present in the state space, and results in files that are notoriously slow to parse, see~\cite{AEKSW22} and \cref{app:explicit-prism}.
In contrast, \UMB offers an efficient, easy-to-use, and widely supported alternative:
Users can describe the transition system on an abstract layer (e.g., programmatically in Python through \UMBI), enabling better integration with other tools, more maintainable and understandable generation scripts (see \href{https://github.com/pmc-tools/umbi/tree/main/examples}{github.com/pmc-tools/umbi/tree/main/examples} for examples), and more effective evaluation pipelines.

\subsubsection{New tool interactions.}
Many PMC tools have a niche where they are best, offering some Pareto optimal point in terms of modelling features, solving time, guarantees provided, and exploiting structural properties of certain models~\cite{BHK+20,ABB+24}.
\UMB allows previously impossible combinations of (i)~tool frontends and backends, and of (ii)~different transformation and analysis algorithms implemented in different tools.
It also allows (iii)~turning PMC tools into components of larger toolchains and portfolios.
We highlight three examples:
\begin{description}[leftmargin=0pt]
\item[\normalfont(i)]\!%
\mcsta is the only PMC tool that supports (Modest and JANI) models with dynamically-sized arrays and user-defined recursive data types.
It is thus used for case studies that can only be succinctly or modularly represented using these features, e.g.~\cite{WBH+26}.
At the same time, \storm's model checking algorithms tend to be faster~\cite{BHK+20}.
Using \UMB, these two strengths can be combined:
Let \mcsta explore the state space and save it to \UMB, which \storm then reads and model-checks---all with little overhead due to the efficiency of \UMB.
\item[\normalfont(ii)]\!%
Vice-versa, for MA, \storm provides the Unif\texttt{+} algorithm~\cite{BHHK15}, while \mcsta also implements the switch-step algorithm~\cite{BF19}, which performs much better on some models~\cite{BHH21}.
Yet \storm offers an efficient bisimulation minimisation for MA~\cite[Chp.~5]{Hen18}. %
Via \UMB, \storm can now generate, minimise, and export the state space of the hard \textit{bitcoin-attack} MA model~\cite{HH19,FC18}, which \mcsta can load and model-check with the switch-step algorithm:\\[0.5pt]
${}$\texttt{storm} \texttt{-{}-jani} \texttt{ba.jani} \texttt{-const} \texttt{MALICIOUS=40,CD=50} \texttt{-{}-janiproperty} \texttt{\textbackslash}\\
${}$\texttt{~~~~~}~\texttt{-e} \texttt{dd-to-sparse} \texttt{-{}-bisimulation} \texttt{-{}-exportbuild} \texttt{ba.umb}\\
${}$\texttt{modest} \texttt{mcsta} \texttt{ba.umb} \texttt{ba.properties.txt} \texttt{-{}-time-bounded-alg} \texttt{SwitchStep}\\[1.5pt]
The \storm command took under 3 seconds to generate the 11475-state state space, minimise it to 7551 states, and write the UMB file on the author's laptop.
\mcsta then checked the model's time-bounded reachability property in 13 seconds.
Trying to check the same property with \storm using Unif\texttt{+}, we aborted when no result was returned after 2 minutes.
The overhead of going through UMB is clearly negligible compared to the model checking time and the time it saves.
\item[\normalfont(iii)]\!%
Using \UMB, we can easily create simple wrappers that call multiple PMC tools on the same state space, explored and exported to \UMB only once, in a portfolio approach to improve performance (as is common in e.g.\ software verification~\cite{BKKR25}).
Prior to \UMB, such an approach would involve one state space exploration---which may take significant time~\cite[Sec.~11]{ABB+24}---in each tool.
Similarly, multiple properties can be checked in parallel; for example, a small, partial state graph can be produced by \stamina{} and then potentially dozens of properties can be checked in parallel by multiple instances of \storm or \prism.
\end{description}

\subsubsection{Rapid prototyping.}
As argued in \cref{sec:math-repr} and~\ref{sec:format}, \UMB is, by design, easily extensible and
can express model classes for which currently only prototypical or even no algorithms exist, such as interval variants of classical models~\cite[Sec.~10]{ABB+24}.
Now, using \UMBI, one can create or parse example models of these classes, and use this interface to develop and compare solution algorithms.
Consequently, \UMB facilitates pushing the frontiers of PMC.
Crucially, researchers working on analysis algorithms can focus on those, processing \UMB files using one of the several parsers available, e.g.\ \UMBI, instead of having to design a modelling language and implement a parser and state space exploration engine.

\section{Format Efficiency}\label{sec:empirical}

\begin{figure}[t]
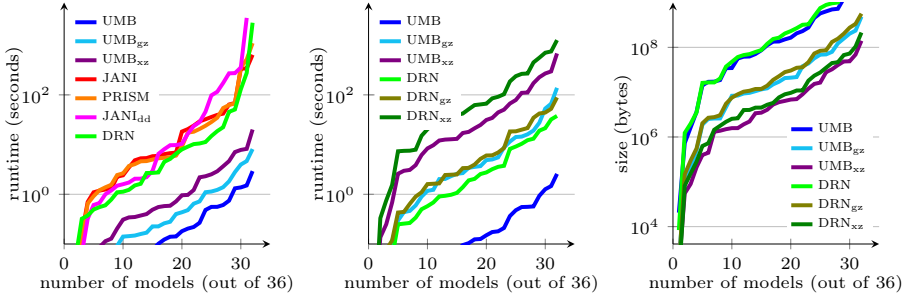

    \pgfplotsset{
      every axis legend/.append style={
        font=\scriptsize,
        inner xsep=2pt,
        inner ysep=1pt,
        row sep=0pt,
      }
    }
    \renewcommand{\quantileplotwidth}{4.5cm}
    \renewcommand{\quantileplotheight}{5cm}
    \begin{minipage}{0.33\textwidth}
    \quantileplot{plotdata/import_time_quantile.csv}
    {
        storm_from-umb_import_sparse/plotblue,
        storm_from-umb-gz_import_sparse/plotcyan,
        storm_from-umb-xz_import_sparse/plotpurple,
        storm_from-jani_import_sparse/plotred,
        storm_from-prism_import_sparse/plotorange,
        storm_from-jani_import_cudd/plotpink,
        storm_from-drn_import_sparse/plotgreen
    }
    {
        UMB,
        UMB$_{\mathrm{gz}}$,
        UMB$_{\mathrm{xz}}$,
        JANI,
        PRISM,
        JANI$_{\mathrm{dd}}$,
        DRN
    }
    {0}{35}{0.1}{7200}{north west}%
    \end{minipage}%
    \begin{minipage}{0.33\textwidth}
    \quantileplot{plotdata/export_time_quantile.csv}
    {
        storm_from-jani_to-umb_sparse/plotblue,
        storm_from-jani_to-umb-gz_sparse/plotcyan,
        storm_from-jani_to-umb-xz_sparse/plotpurple,
        storm_from-jani_to-drn_sparse/plotgreen,
        storm_from-jani_to-drn-gz_sparse/plotgreenyellow,
        storm_from-jani_to-drn-xz_sparse/plotdarkgreen
    }
    {
        UMB,
        UMB$_{\mathrm{gz}}$,
        UMB$_{\mathrm{xz}}$,
        DRN,
        DRN$_{\mathrm{gz}}$,
        DRN$_{\mathrm{xz}}$,
    }
    {0}{35}{0.1}{7200}{north west}%
    \end{minipage}%
    \begin{minipage}{0.33\textwidth}%
    \quantileplotsize{plotdata/export_size_quantile.csv}
    {
        storm_from-jani_to-umb_sparse/plotblue,
        storm_from-jani_to-umb-gz_sparse/plotcyan,
        storm_from-jani_to-umb-xz_sparse/plotpurple,
        storm_from-jani_to-drn_sparse/plotgreen,
        storm_from-jani_to-drn-gz_sparse/plotgreenyellow,
        storm_from-jani_to-drn-xz_sparse/plotdarkgreen
    }
    {
        UMB,
        UMB$_{\mathrm{gz}}$,
        UMB$_{\mathrm{xz}}$,
        DRN,
        DRN$_{\mathrm{gz}}$,
        DRN$_{\mathrm{xz}}$,
    }
    {0}{35}{4096}{1e9}{south east}
    \end{minipage}%
    \caption{
    Summary of our format efficiency experiments.
    From left to right, we report (i)~the time to obtain an explicit-state representation for various model formats, (ii)~the export time for explicit-state formats, and (iii)~the file size of explicit-state formats, also including compression.
    }
    \label{fig:empirical}
\end{figure}

\noindent We evaluate \UMB in terms of reading/writing speed and overall size on 36 standard DTMC, CTMC, and MDP benchmark models from~\cite{HKPQR19}.
\cref{app:experiments} provides more details on the experimental setup as well as results for \prism and \mcsta, which qualitatively agree with those of \storm described here.

High-level modelling formalisms like JANI and the \prism\ language usually require explicit-state representations of models to be constructed in-memory before probabilistic model checking can be performed.
\cref{fig:empirical} (left) shows that loading a \UMB file is significantly faster than generating such explicit representations from either \prism\ or JANI models or loading from an explicit-state DRN file. %

To compare with other explicit-state formats, we evaluate size and time to export to DRN and \UMB.
In \cref{fig:empirical} (middle) we show the runtime to export a model.
Here, (uncompressed) \UMB is exported consistently significantly faster than (uncompressed) DRN (note the log-scale).
As expected, compression significantly slows down both formats, and they become more closely aligned (with \UMB still ahead).
For size, \cref{fig:empirical} (right) shows that DRN and \UMB{} perform very comparable, and compression saves a similar amount of space.
Aligning with usual observations, XZ compression is better but slower than GZip.

In summary, \UMB is not only the most flexible format, but also concise and by far the fastest way of loading explicit-state models.

%% file: 9_conclusion.tex
\section{Conclusion}
We presented the unified Markov binary (UMB) format,
a flexible, efficient, and extensible file format for explicit-state probabilistic models,
backed by extensive tool support.
Possible future extensions include support for
models with symbolic expressions,
storage of policies/strategies,
and property specification.

\subsubsection*{Data availability statement.}
Our Git repositories at \href{https://github.com/pmc-tools/}{github.com/pmc-tools} provide access to the software and specifications discussed in this paper.
Full tables of our benchmark results are available at \href{https://pmc-tools.github.io/umb-benchmarks/}{pmc-tools.github.io/umb-benchmarks}.

\subsubsection*{Funding.}
This work was partly supported by
the European Union's Horizon 2020 research and innovation programme under Marie Skłodowska-Curie grant agreement 101008233 (MISSION),
the Interreg North Sea project STORM\_SAFE,
the KI-Starter Project ``Verifying AI Systems under Partial Observability'' of the Ministry of Culture and Science of the German State of North Rhine-Westphalia, %
and NWO VIDI grant VI.Vidi.223.110 (TruSTy).
Experiments were performed with computing resources granted by RWTH Aachen University under project rwth1632. %

%% file: 10_appx_umbt.tex
\section{Example UMB File (textual version)}\label{app:example}

Below, we give an illustration of the contents of an example {\tt .umb} file,
with the binary files converted to text for readability.
This corresponds to the example MDP in Figure~\ref{fig:prelim-example} (\cref{sec:math-repr}),
with the addition of:
(i) an atomic proposition $g$ that is true only in state $s_3$;
(ii) a reward structure $r$ that assigns 1 to state-action pair $(s_0,b)$;
(iii) state valuations for each state using two variables {\tt x} and {\tt b}.

\begin{multicols}{2}
\begin{lstlisting}[basicstyle=\ttfamily\scriptsize]
/index.json:
{
  "format-version": 1,
  "format-revision": 0,
  "model-data": {},
  "file-data": {
    "tool": "PRISM",
    "tool-version": "4.10",
    "creation-date": 1769360247
  },
  "transition-system": {
    "time": "discrete",
    "#players": 1,
    "#states": 4,
    "#initial-states": 1,
    "#choices": 5,
    "#choice-actions": 2,
    "#branches": 8,
    "#branch-actions": 0,
    "#observations": 0,
    "branch-probability-type": {
      "type": "double",
      "size": 64
    }
  },
  "annotations": {
    "aps": {
      "g": {
        "alias": "g",
        "applies-to": [
          "states"
        ],
        "type": {
          "type": "bool",
          "size": 1
        }
      }
    },
    "rewards": {
      "r": {
        "alias": "r",
        "applies-to": [
          "choices"
        ],
        "type": {
          "type": "double",
          "size": 64
        }
      }
    }
  },
  ...
\end{lstlisting}
\begin{lstlisting}[basicstyle=\ttfamily\scriptsize]
  ...
  "valuations": {
    "states": {
      "unique": true,
      "classes": [
        {
          "variables": [
            {
              "name": "x",
              "type": {
                "type": "uint",
                "size": 2
              }
            },
            {
              "name": "b",
              "type": {
                "type": "bool",
                "size": 1
              }
            },
            {
              "padding": 5
            }
          ]
        }
      ]
    }
  }
}
/state-to-choices.bin:
[0,2,3,4,5]
/choice-to-branches.bin:
[0,2,4,6,7,8]
/branch-to-probability.bin:
[0.5,0.5,0.2,0.8,0.9,0.1,1.0,1.0]
/branch-to-target.bin:
[1,2,2,3,0,1,2,3]
/state-is-initial.bin:
[1000]
/actions/choices/string-mapping.bin:
[0,1,2]
/actions/choices/strings.bin:
[a,b]
/actions/choices/values.bin:
[0,1,0,0,0]
/annotations/aps/g/states/values.bin:
[0001]
/annotations/rewards/r/choices/values.bin:
[0.0,1.0,0.0,0.0,0.0]
/valuations/states/valuations.bin:
[.....|0|00,.....|0|01,.....|0|10,.....|1|10]
\end{lstlisting}
\end{multicols}
\clearpage %

%% file: 10_appx_models.tex
\section{Further Details on Model Encodings}\label{app:ats}

\cref{tab:atsmodels}, below, gives a brief summary of how the various model classes discussed
in \cref{sec:math-repr} make use of the different entities and annotations that
are used in an annotated transition system (ATS).
Note also the correspondence between these features
and the fields of the {\tt "transition-system"} object
in the JSON-encoded metadata for the earlier example file,
which defines model type.

\begin{table}[!h]
\vspace*{-1em}
\setlength{\tabcolsep}{3pt}
\centering
\begin{tabular}{lccccccc}
Model type                   & \rotatebox{90}{States}     & \rotatebox{90}{Choices}    & \rotatebox{90}{Branches}   & \rotatebox{90}{\#Players}    & \rotatebox{90}{Observations} & \rotatebox{90}{\parbox{2cm}{Branch \\ probabilities}} & \rotatebox{90}{Exit rates}
\\
\midrule
Markov decision process      & \checkmark & \checkmark & \checkmark & 1  & -          & \checkmark & -          \\
Discrete-time Markov chain   & \checkmark & $\one$     & \checkmark & 0  & -          & \checkmark & -          \\
Continuous-time Markov chain & \checkmark & $\one$     & \checkmark & 0  & -          & \checkmark & \checkmark \\
Markov automaton             & \checkmark & \checkmark & \checkmark & 1  & -          & \checkmark$^2$ & \checkmark \\
Partially observable MDP     & \checkmark & \checkmark & \checkmark & 1  & \checkmark & \checkmark & -          \\
Interval Markov chain        & \checkmark & $\one$     & \checkmark & 0  & -          & Intervals & -          \\
Interval MDP                 & \checkmark & \checkmark & \checkmark & 1  & -          & Intervals  & -          \\
Turn-based stochastic game   & \checkmark & \checkmark & \checkmark & >1 & -          & \checkmark & -          \\
Labelled transition system   & \checkmark & \checkmark & $\one$     & 1  & -          & -          & -          \\
\midrule
\end{tabular}
\vspace*{1em}
\caption{
    Concise summary of how established model types are mapped to our ATS formalism.
    Each row corresponds to a model type and the columns describe which concepts are used.
    Specifically, \checkmark{} indicates usage, $\one$ indicates trivial/singleton usage, and \checkmark$^2$ marks double usage (specifically for probabilities and rates).
}
\label{tab:atsmodels}
\end{table}

%% file: 10_appx_experiments.tex
\section{Details on Experiments}\label{app:experiments}

\cref{app:explicit-prism} showcases that the common practice of encoding explicit-state models in the \lang{PRISM} language (see paragraph \enquote{Explicit-state model communication} in \cref{sec:tooldemo}) leads to enormous runtimes.
Then, \cref{app:exp-setup} and \cref{app:exp-plots} concern the evaluation of format efficiency in \cref{sec:empirical}, first describing the setup and then providing the full results.

\subsection{Encoding transition systems explicitly in high-level languages.}\label{app:explicit-prism}

    In this experiment, we investigate parsing times for transition systems that describe an explicit-state encoding in a high-level language such as \prism. To this end, we generated a random transition system using \href{https://github.com/pmc-tools/umbi/tree/main/umbi/examples/ats/random_game.py}{\umbi} and \href{https://github.com/pmc-tools/umbi/tree/main/examples/ats_smg_to_prism.py}{encoded} it as a flat \prism file. We report that while transition systems with 10k states can be parsed by \prism in a matter of seconds, models with only 100k states cannot be parsed within an hour. 
    We stress that in this case, long parsing times are not a limitation of the tool---\prism can parse \emph{compact} encodings of models with tens of millions of states in a matter of seconds---but rather a consequence of the explicit encoding format, which generates prohibitively large files and necessitates linear parsing of unstructured state transitions with extensive lookup operations.

\subsection{Experimental Setup for \cref{sec:empirical}}\label{app:exp-setup}

\paragraph{Tools.}
We consider the three tools that natively support \UMB:
\begin{itemize}
    \item 
\prism version \texttt{4.10.2-dev 8a05345} \url{https://www.prismmodelchecker.org}
\item \storm version \texttt{1.12.0 cdd37bc} \url{https://www.stormchecker.org}
\item \mcsta version \texttt{v3.1.311-g14c460466} \url{https://www.modestchecker.net}
\end{itemize}
For all of them, we compare the time to parse a case study given in the \UMB format to the time it takes to parse (or explore) it in other formats the tool supports, namely the high-level languages \jani~\cite{BDH+17} and the \lang{PRISM} language~\cite{prismlang}, as well as the explicit textual encoding \texttt{.drn}~\cite{drn} for \storm and \texttt{.tra} for \prism.
We also measure the time it takes to export a model and the resulting file sizes---where we additionally consider the textual encodings \texttt{.mrm} (from the IMCA format~\cite{GHKN12}) and \texttt{.aut} (from CADP~\cite{GLMS13})
for \tool{mcsta}.

\paragraph{Benchmarks.}
We consider all QComp 2019 and 2020~\cite{HHH+19,BHK+20} benchmarks and select 36 DTMC, CTMC, and MDP benchmark models that are available in the PRISM language as these are the model types that all three tools support.

\begin{figure}[tpb]
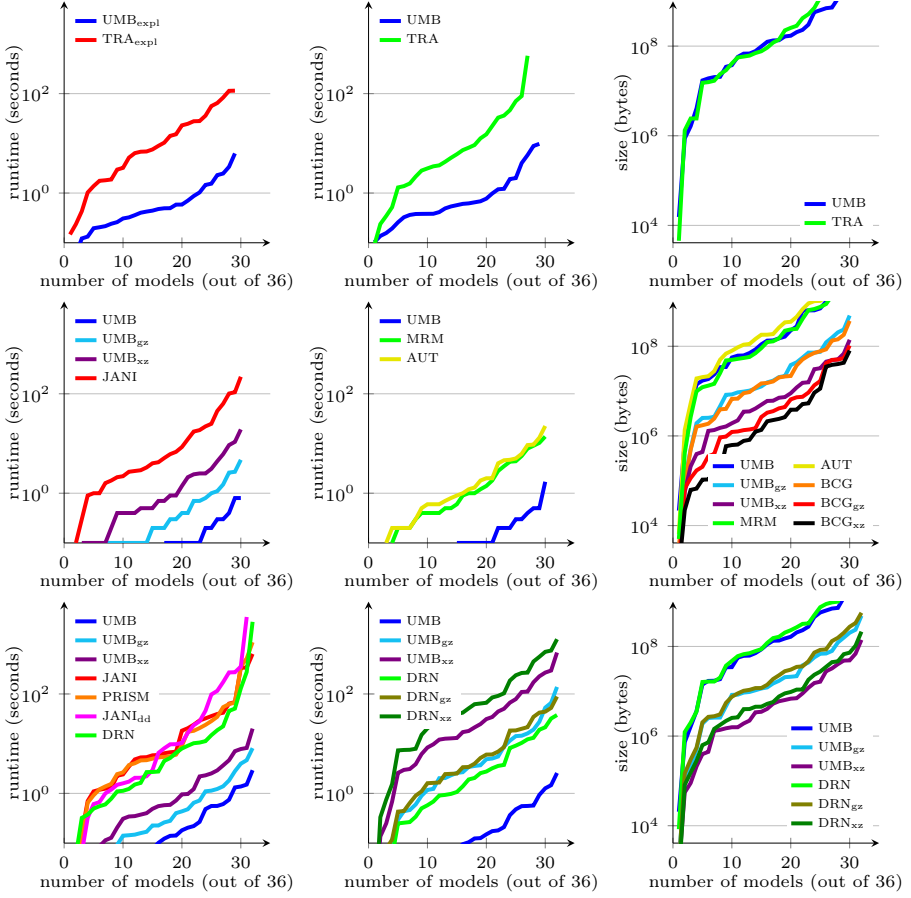

    \pgfplotsset{
      every axis legend/.append style={
        font=\scriptsize,
        inner xsep=2pt,
        inner ysep=1pt,
        row sep=0pt,
      }
    }
    \renewcommand{\quantileplotwidth}{4.5cm}
    \renewcommand{\quantileplotheight}{5cm}
    \begin{minipage}{0.33\textwidth}
    \quantileplot{plotdata/import_time_quantile.csv}
    {
        prism_from-umb_import_ex/plotblue,
        prism_from-tra_import_ex/plotgreen,
        prism_from-prism_import_ex/plotred
    }
    {
        UMB$_\mathrm{expl}$,
        TRA$_\mathrm{expl}$,
        PRISM$_\mathrm{expl}$,
    }
    {0}{35}{0.1}{7200}{north west}%
    \end{minipage}%
    \begin{minipage}{0.33\textwidth}
    \quantileplot{plotdata/export_time_quantile.csv}
    {
        prism_from-prism_to-umb_ex/plotblue,
        prism_from-prism_to-tra_ex/plotgreen
    }
    {
        UMB,
        TRA
    }
    {0}{35}{0.1}{7200}{north west}%
    \end{minipage}%
    \begin{minipage}{0.33\textwidth}%
    \quantileplotsize{plotdata/export_size_quantile.csv}
    {
        prism_from-prism_to-umb_ex/plotblue,
        prism_from-prism_to-tra_ex/plotgreen
    }
    {
        UMB,
        TRA
    }
    {0}{35}{4096}{1e9}{south east}
    \end{minipage}\\
       \begin{minipage}{0.33\textwidth}
    \quantileplot{plotdata/import_time_quantile.csv}
    {
        modest_from-umb_import_default/plotblue,
        modest_from-umb-gz_import_default/plotcyan,
        modest_from-umb-xz_import_default/plotpurple,
        modest_from-jani_import_default/plotred
    }
    {
        UMB,
        UMB$_{\mathrm{gz}}$,
        UMB$_{\mathrm{xz}}$,
        JANI
    }
    {0}{35}{0.1}{7200}{north west}%
    \end{minipage}%
    \begin{minipage}{0.33\textwidth}
    \quantileplot{plotdata/export_time_quantile.csv}
    {
        modest_from-jani_to-umb_default/plotblue,
        modest_from-jani_to-imca_default/plotgreen,
        modest_from-jani_to-aut_default/plotyellow
    }
    {
        UMB,
        MRM,
        AUT
    }
    {0}{35}{0.1}{7200}{north west}%
    \end{minipage}%
    \begin{minipage}{0.33\textwidth}%
	\renewcommand{\quantileplotlegendcols}{2}
    \quantileplotsize{plotdata/export_size_quantile.csv}
    {
        modest_from-jani_to-umb_default/plotblue,
        modest_from-jani_to-aut_default/plotyellow,
        modest_from-jani_to-umb-gz_default/plotcyan,
        modest_bcgio_aut-to-bcg_default/plotorange,
        modest_from-jani_to-umb-xz_default/plotpurple,
        modest_bcgio_aut-to-bcg-gz_default/plotred,
        modest_from-jani_to-imca_default/plotgreen,
        modest_bcgio_aut-to-bcg-xz_default/black
    }
    {
        UMB,
        AUT,
        UMB$_{\mathrm{gz}}$,
        BCG,
        UMB$_{\mathrm{xz}}$,
        BCG$_{\mathrm{gz}}$,
        MRM,
        BCG$_{\mathrm{xz}}$
    }
    {0}{35}{4096}{1e9}{south east}
    \end{minipage}\\
       \begin{minipage}{0.33\textwidth}
    \quantileplot{plotdata/import_time_quantile.csv}
    {
        storm_from-umb_import_sparse/plotblue,
        storm_from-umb-gz_import_sparse/plotcyan,
        storm_from-umb-xz_import_sparse/plotpurple,
        storm_from-jani_import_sparse/plotred,
        storm_from-prism_import_sparse/plotorange,
        storm_from-jani_import_cudd/plotpink,
        storm_from-drn_import_sparse/plotgreen
    }
    {
        UMB,
        UMB$_{\mathrm{gz}}$,
        UMB$_{\mathrm{xz}}$,
        JANI,
        PRISM,
        JANI$_{\mathrm{dd}}$,
        DRN
    }
    {0}{35}{0.1}{7200}{north west}%
    \end{minipage}%
    \begin{minipage}{0.33\textwidth}
    \quantileplot{plotdata/export_time_quantile.csv}
    {
        storm_from-jani_to-umb_sparse/plotblue,
        storm_from-jani_to-umb-gz_sparse/plotcyan,
        storm_from-jani_to-umb-xz_sparse/plotpurple,
        storm_from-jani_to-drn_sparse/plotgreen,
        storm_from-jani_to-drn-gz_sparse/plotgreenyellow,
        storm_from-jani_to-drn-xz_sparse/plotdarkgreen
    }
    {
        UMB,
        UMB$_{\mathrm{gz}}$,
        UMB$_{\mathrm{xz}}$,
        DRN,
        DRN$_{\mathrm{gz}}$,
        DRN$_{\mathrm{xz}}$,
    }
    {0}{35}{0.1}{7200}{north west}%
    \end{minipage}%
    \begin{minipage}{0.33\textwidth}%
    \quantileplotsize{plotdata/export_size_quantile.csv}
    {
        storm_from-jani_to-umb_sparse/plotblue,
        storm_from-jani_to-umb-gz_sparse/plotcyan,
        storm_from-jani_to-umb-xz_sparse/plotpurple,
        storm_from-jani_to-drn_sparse/plotgreen,
        storm_from-jani_to-drn-gz_sparse/plotgreenyellow,
        storm_from-jani_to-drn-xz_sparse/plotdarkgreen
    }
    {
        UMB,
        UMB$_{\mathrm{gz}}$,
        UMB$_{\mathrm{xz}}$,
        DRN,
        DRN$_{\mathrm{gz}}$,
        DRN$_{\mathrm{xz}}$,
    }
    {0}{35}{4096}{1e9}{south east}
    \end{minipage}%
    \caption{
    Summary of our format efficiency experiments for \tool{Prism} (top), \tool{mcsta} (middle), and \tool{Storm} (bottom).
    From left to right, we report (i)~the time to obtain an explicit-state representation for various model formats, (ii)~the export time for explicit-state formats, and (iii)~the file size of explicit-state formats, also including compression.
    The plots in the bottom row coincide with \Cref{fig:empirical}.
    }
    \label{app:fig:empirical}
    \end{figure}

     \begin{figure}[tpb]
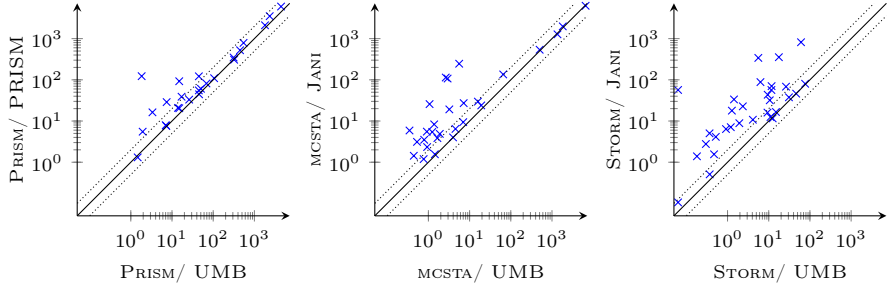

         \centering
 \scatterplot{plotdata/full_time_median.csv}{prism_from-umb_check_ex}{\prism / \UMB}{prism_from-prism_check_ex}{\prism / \lang{PRISM}}{false}%
  \scatterplot{plotdata/full_time_median.csv}{modest_from-umb_check_default}{\mcsta / \UMB}{modest_from-jani_check_default}{\mcsta / \jani}{false}%
   \scatterplot{plotdata/full_time_median.csv}{storm_from-umb_check_sparse}{\storm / \UMB}{storm_from-jani_check_sparse}{\storm / \jani}{false}
\caption{Comparison of model checking time of \prism (left), \mcsta (middle), and \storm (right) when using \UMB vs. a high-level language. Timings in seconds.}
         \label{fig:scatter}
     \end{figure}
\paragraph{Quantile plots.}
We present our data as quantile plots in \Cref{app:fig:empirical}.
In these, we have one line per tool $T$.
A point $(x, y)$ for tool T indicates that the $x$ smallest/fastest benchmark of T have size/runtime at most $y$.

\paragraph{Execution environment.}
All executions were run on the Intel Xeon 8468 Sapphire systems running 64-bit Rocky Linux 9.6.
The timeout for each benchmark was 7200 seconds.
Executions were limited to 4 CPU cores and 32 GB of memory using the \tool{Slurm} batch system.
Files were read from and written to an NVMe SSD disk.
Each experiment is repeated 5 times and we consider the median measurement.

\subsection{Full Experimental Results for \cref{sec:empirical}}\label{app:exp-plots}
\cref{app:fig:empirical} shows the experimental results for all 3 tools, while \cref{fig:empirical} only provides the results for \storm (i.e., the bottom row of \cref{app:fig:empirical}).
The general trend reported in \cref{sec:empirical} is visible for all tools: \UMB is significantly faster to import and export, as well as comparably concise.

We also measure the runtime of the entire model checking pipeline, i.e., the time it takes to obtain an explicit model representation plus the time it takes to verify a property on that model.
 \Cref{fig:scatter} compares the time for \prism (left), \mcsta (middle), and \storm (right) when given a \UMB model as input (x-axis) vs. a \lang{PRISM}/\jani model (y-axis).
We observe that model checking for \UMB input is consistently faster compared to high-level inputs across all tools, sometimes by several orders of magnitude.

%% file: references.bib
@misc{drn,
  key          = {_zzz_www_stormchecker_org_doc},
  author       = {{Storm developers}},
  title        = {{DRN} format description},
  url          = {https://www.stormchecker.org/documentation/background/drn.html}
}

@misc{prismlang,
  key          = {_zzz_www_prismmodelchecker_org_manual_t},
  author       = {{PRISM developers}},
  title        = {{PRISM} Modelling Language Documentation},
  url          = {https://www.prismmodelchecker.org/manual/ThePRISMLanguage}
}

@misc{tra,
  key          = {_zzz_www_prismmodelchecker_org_manual_1},
  author       = {{PRISM developers}},
  title        = {{PRISM} Explicit Model Format Specification},
  url          = {https://www.prismmodelchecker.org/manual/Appendices/ExplicitModelFiles}
}

@article{HJQW26,
  author       = {Arnd Hartmanns and
                  Sebastian Junges and
                  Tim Quatmann and
                  Maximilian Weininger},
  title        = {The Revised Practitioner's Guide to {MDP} Model Checking Algorithms},
  journal      = {Int. J. Softw. Tools Technol. Transf.},
  year         = {2026},
  doi          = {10.1007/s10009-026-00848-y}
}

@inproceedings{WBH+26,
  author       = {Nick Waddoups and
                  Jonah Boe and
                  Arnd Hartmanns and
                  Prabal Basu and
                  Sanghamitra Roy and
                  Koushik Chakraborty and
                  Zhen Zhang},
  title        = {Probabilistic Verification for Modular Network-on-Chip Systems},
  booktitle    = {27th International Conference on Verification, Model Checking, and Abstract Interpretation ({VMCAI} 2026)},
  series       = {Lecture Notes in Computer Science},
  publisher    = {Springer},
  year         = {2026},
  doi          = {10.1007/978-3-032-15700-3\_18}
}

@article{BKKR25,
  author       = {Dirk Beyer and
                  Sudeep Kanav and
                  Tobias Kleinert and
                  Cedric Richter},
  title        = {Construction of verifier combinations from off-the-shelf components},
  journal      = {Formal Methods Syst. Des.},
  volume       = {66},
  number       = {1},
  pages        = {99--130},
  year         = {2025},
  doi          = {10.1007/S10703-024-00449-Y}
}

@techreport{LMHVJ25,
  author       = {Pim Leerkes and
                  Ivo Melse and
                  Linus Heck and
                  Matthias Volk and
                  Sebastian Junges},
  title        = {{S}tormvogel: Probabilistic Model Checking for Almost Everyone},
  year         = {2025},
  institution  = {Radboud University Nijmegen},
  url          = {https://hdl.handle.net/2066/325553}
}

@inproceedings{KNP25,
  author       = {Marta Kwiatkowska and
                  Gethin Norman and
                  David Parker},
  editor       = {Nathalie Bertrand and
                  Clemens Dubslaff and
                  Sascha Kl{\"{u}}ppelholz},
  title        = {Probabilistic Model Checking: Applications and Trends},
  booktitle    = {Principles of Formal Quantitative Analysis},
  series       = {Lecture Notes in Computer Science},
  volume       = {15760},
  pages        = {158--173},
  publisher    = {Springer},
  year         = {2025},
  doi          = {10.1007/978-3-031-97439-7\_7}
}

@incollection{ABB+24,
  author       = {Roman Andriushchenko and
                  Alexander Bork and
                  Carlos E. Budde and
                  Milan \v{C}e\v{s}ka and
                  Ernst Moritz Hahn and
                  Arnd Hartmanns and
                  Bryant Israelsen and
                  Nils Jansen and
                  Joshua Jeppson and
                  Sebastian Junges and
                  Maximilian A. K\"ohl and
                  Bettina K\"onighofer and
                  Jan K\v{r}et\'insk\'y and
                  Tobias Meggendorfer and
                  David Parker and
                  Stefan Pranger and
                  Tim Quatmann and
                  Enno Ruijters and
                  Landon Taylor and
                  Matthias Volk and
                  Maximilian Weininger and
                  Zhen Zhang},
  editor       = {Dirk Beyer and
                  Arnd Hartmanns and
                  Fabrice Kordon},
  title        = {Tools at the Frontiers of Quantitative Verification: {QC}omp 2023 Competition Report},
  booktitle    = {TOOLympics Challenge 2023 -- Updates, Results, Successes of the Formal-Methods Competitions ({TOOLympics} 2023)},
  pages        = {90--146},
  year         = {2024},
  publisher    = {Springer},
  doi          = {10.1007/978-3-031-67695-6\_4}
}

@inproceedings{HFRSL24,
  author       = {Qi Heng Ho and
                  Martin S. Feather and
                  Federico Rossi and
                  Zachary Sunberg and
                  Morteza Lahijanian},
  editor       = {Negar Kiyavash and
                  Joris M. Mooij},
  title        = {Sound Heuristic Search Value Iteration for Undiscounted {POMDP}s with Reachability Objectives},
  booktitle    = {40th Conference on Uncertainty in Artificial Intelligence ({UAI} 2024)},
  series       = {Proceedings of Machine Learning Research},
  volume       = {244},
  pages        = {1681--1697},
  publisher    = {{PMLR}},
  year         = {2024},
  url          = {https://proceedings.mlr.press/v244/ho24b.html}
}

@inproceedings{JND+24,
  author       = {Chris Johannsen and
                  Karthik Nukala and
                  Rohit Dureja and
                  Ahmed Irfan and
                  Natarajan Shankar and
                  Cesare Tinelli and
                  Moshe Y. Vardi and
                  Kristin Yvonne Rozier},
  editor       = {Arie Gurfinkel and
                  Vijay Ganesh},
  title        = {The {MoXI} Model Exchange Tool Suite},
  booktitle    = {36th International Conference on Computer Aided Verification ({CAV} 2024)},
  series       = {Lecture Notes in Computer Science},
  volume       = {14681},
  pages        = {203--218},
  publisher    = {Springer},
  year         = {2024},
  doi          = {10.1007/978-3-031-65627-9\_10}
}

@inproceedings{MW24,
  author       = {Tobias Meggendorfer and
                  Maximilian Weininger},
  editor       = {Arie Gurfinkel and
                  Vijay Ganesh},
  title        = {Playing Games with Your {PET:} Extending the Partial Exploration Tool to Stochastic Games},
  booktitle    = {36th International Conference on Computer Aided Verification ({CAV} 2024)},
  series       = {Lecture Notes in Computer Science},
  volume       = {14683},
  pages        = {359--372},
  publisher    = {Springer},
  year         = {2024},
  doi          = {10.1007/978-3-031-65633-0\_16}
}

@inproceedings{JVI+23,
  author       = {Joshua Jeppson and
                  Matthias Volk and
                  Bryant Israelsen and
                  Riley Roberts and
                  Andrew Williams and
                  Lukas Buecherl and
                  Chris J. Myers and
                  Hao Zheng and
                  Chris Winstead and
                  Zhen Zhang},
  editor       = {Nils Jansen and
                  Mirco Tribastone},
  title        = {{STAMINA} in {C++:} Modernizing an Infinite-State Probabilistic Model Checker},
  booktitle    = {20th International Conference on Quantitative Evaluation of Systems ({QEST} 2023)},
  series       = {Lecture Notes in Computer Science},
  volume       = {14287},
  pages        = {101--109},
  publisher    = {Springer},
  year         = {2023},
  doi          = {10.1007/978-3-031-43835-6\_7}
}

@inproceedings{AEKSW22,
  author       = {Muqsit Azeem and
                  Alexandros Evangelidis and
                  Jan Kret{\'{\i}}nsk{\'{y}} and
                  Alexander Slivinskiy and
                  Maximilian Weininger},
  editor       = {Ahmed Bouajjani and
                  Luk{\'{a}}s Hol{\'{\i}}k and
                  Zhilin Wu},
  title        = {Optimistic and Topological Value Iteration for Simple Stochastic Games},
  booktitle    = {20th International Symposium on Automated Technology for Verification and Analysis ({ATVA} 2022)},
  series       = {Lecture Notes in Computer Science},
  volume       = {13505},
  pages        = {285--302},
  publisher    = {Springer},
  year         = {2022},
  doi          = {10.1007/978-3-031-19992-9\_18}
}

@article{HJKQV22,
  author       = {Christian Hensel and
                  Sebastian Junges and
                  Joost-Pieter Katoen and
                  Tim Quatmann and
                  Matthias Volk},
  title        = {The probabilistic model checker {S}torm},
  journal      = {Int. J. Softw. Tools Technol. Transf.},
  volume       = {24},
  number       = {4},
  pages        = {589--610},
  year         = {2022},
  doi          = {10.1007/S10009-021-00633-Z}
}

@inproceedings{ACJKS21,
  author       = {Roman Andriushchenko and
                  Milan Ceska and
                  Sebastian Junges and
                  Joost-Pieter Katoen and
                  Simon Stupinsk{\'{y}}},
  editor       = {Alexandra Silva and
                  K. Rustan M. Leino},
  title        = {{PAYNT}: A Tool for Inductive Synthesis of Probabilistic Programs},
  booktitle    = {33rd International Conference on Computer Aided Verification ({CAV} 2021)},
  series       = {Lecture Notes in Computer Science},
  volume       = {12759},
  pages        = {856--869},
  publisher    = {Springer},
  year         = {2021},
  doi          = {10.1007/978-3-030-81685-8\_40}
}

@inproceedings{AJK+21,
  author       = {Pranav Ashok and
                  Mathias Jackermeier and
                  Jan Kret{\'{\i}}nsk{\'{y}} and
                  Christoph Weinhuber and
                  Maximilian Weininger and
                  Mayank Yadav},
  editor       = {Jan Friso Groote and
                  Kim Guldstrand Larsen},
  title        = {{dtControl} 2.0: Explainable Strategy Representation via Decision Tree Learning Steered by Experts},
  booktitle    = {27th International Conference on Tools and Algorithms for the Construction and Analysis of Systems ({TACAS} 2021)},
  series       = {Lecture Notes in Computer Science},
  volume       = {12652},
  pages        = {326--345},
  publisher    = {Springer},
  year         = {2021},
  doi          = {10.1007/978-3-030-72013-1\_17}
}

@article{BHH21,
  author       = {Yuliya Butkova and
                  Arnd Hartmanns and
                  Holger Hermanns},
  title        = {A {M}odest Approach to {M}arkov Automata},
  journal      = {{ACM} Trans. Model. Comput. Simul.},
  volume       = {31},
  number       = {3},
  pages        = {14:1--14:34},
  year         = {2021},
  doi          = {10.1145/3449355}
}

@inproceedings{BHK+20,
  author       = {Carlos E. Budde and
                  Arnd Hartmanns and
                  Michaela Klauck and
                  Jan Kret{\'{\i}}nsk{\'{y}} and
                  David Parker and
                  Tim Quatmann and
                  Andrea Turrini and
                  Zhen Zhang},
  editor       = {Tiziana Margaria and
                  Bernhard Steffen},
  title        = {On Correctness, Precision, and Performance in Quantitative Verification - {QComp} 2020 Competition Report},
  booktitle    = {9th International Symposium on Leveraging Applications of Formal Methods, Verification and Validation ({ISoLA} 2020)},
  series       = {Lecture Notes in Computer Science},
  volume       = {12479},
  pages        = {216--241},
  publisher    = {Springer},
  year         = {2020},
  doi          = {10.1007/978-3-030-83723-5\_15}
}

@inproceedings{GHHKS20,
  author       = {Timo P. Gros and
                  Holger Hermanns and
                  J{\"{o}}rg Hoffmann and
                  Michaela Klauck and
                  Marcel Steinmetz},
  editor       = {Alexey Gotsman and
                  Ana Sokolova},
  title        = {Deep Statistical Model Checking},
  booktitle    = {40th {IFIP} {WG} 6.1 International Conference on Formal Techniques for Distributed Objects, Components, and Systems ({FORTE} 2020)},
  series       = {Lecture Notes in Computer Science},
  volume       = {12136},
  pages        = {96--114},
  publisher    = {Springer},
  year         = {2020},
  doi          = {10.1007/978-3-030-50086-3\_6}
}

@inproceedings{BF19,
  author       = {Yuliya Butkova and
                  Gereon Fox},
  editor       = {Tom{\'{a}}s Vojnar and
                  Lijun Zhang},
  title        = {Optimal Time-Bounded Reachability Analysis for Concurrent Systems},
  booktitle    = {25th International Conference on Tools and Algorithms for the Construction and Analysis of Systems ({TACAS} 2019)},
  series       = {Lecture Notes in Computer Science},
  volume       = {11428},
  pages        = {191--208},
  publisher    = {Springer},
  year         = {2019},
  doi          = {10.1007/978-3-030-17465-1\_11}
}

@inproceedings{HHH+19,
  author       = {Ernst Moritz Hahn and
                  Arnd Hartmanns and
                  Christian Hensel and
                  Michaela Klauck and
                  Joachim Klein and
                  Jan Kret{\'{\i}}nsk{\'{y}} and
                  David Parker and
                  Tim Quatmann and
                  Enno Ruijters and
                  Marcel Steinmetz},
  editor       = {Tom{\'{a}}s Vojnar and
                  Lijun Zhang},
  title        = {The 2019 Comparison of Tools for the Analysis of Quantitative Formal Models - ({QComp} 2019 Competition Report)},
  booktitle    = {25th International Conference on Tools and Algorithms for the Construction and Analysis of Systems ({TACAS} 2019)},
  series       = {Lecture Notes in Computer Science},
  volume       = {11429},
  pages        = {69--92},
  publisher    = {Springer},
  year         = {2019},
  doi          = {10.1007/978-3-030-17502-3\_5}
}

@inproceedings{HH19,
  author       = {Arnd Hartmanns and
                  Holger Hermanns},
  editor       = {Markus Kr{\"{o}}tzsch and
                  Daria Stepanova},
  title        = {A {M}odest {M}arkov Automata Tutorial},
  booktitle    = {Reasoning Web -- 15th International Summer School 2019, Tutorial Lectures},
  series       = {Lecture Notes in Computer Science},
  volume       = {11810},
  pages        = {250--276},
  publisher    = {Springer},
  year         = {2019},
  doi          = {10.1007/978-3-030-31423-1\_8}
}

@inproceedings{HKPQR19,
  author       = {Arnd Hartmanns and
                  Michaela Klauck and
                  David Parker and
                  Tim Quatmann and
                  Enno Ruijters},
  editor       = {Tom{\'{a}}s Vojnar and
                  Lijun Zhang},
  title        = {The Quantitative Verification Benchmark Set},
  booktitle    = {25th International Conference on Tools and Algorithms for the Construction and Analysis of Systems ({TACAS} 2019)},
  series       = {Lecture Notes in Computer Science},
  volume       = {11427},
  pages        = {344--350},
  publisher    = {Springer},
  year         = {2019},
  doi          = {10.1007/978-3-030-17462-0\_20}
}

@incollection{BAFK18,
  author       = {Christel Baier and
                  Luca de Alfaro and
                  Vojtech Forejt and
                  Marta Kwiatkowska},
  editor       = {Edmund M. Clarke and
                  Thomas A. Henzinger and
                  Helmut Veith and
                  Roderick Bloem},
  title        = {Model Checking Probabilistic Systems},
  booktitle    = {Handbook of Model Checking},
  pages        = {963--999},
  publisher    = {Springer},
  year         = {2018},
  doi          = {10.1007/978-3-319-10575-8\_28}
}

@inproceedings{FC18,
  author       = {Ansgar Fehnker and
                  Kaylash Chaudhary},
  editor       = {Aaron Dutle and
                  C{\'{e}}sar A. Mu{\~{n}}oz and
                  Anthony Narkawicz},
  title        = {Twenty Percent and a Few Days -- Optimising a {B}itcoin Majority Attack},
  booktitle    = {10th International {NASA} Formal Methods Symposium ({NFM} 2018)},
  series       = {Lecture Notes in Computer Science},
  volume       = {10811},
  pages        = {157--163},
  publisher    = {Springer},
  year         = {2018},
  doi          = {10.1007/978-3-319-77935-5\_11}
}

@phdthesis{Hen18,
  author       = {Christian Hensel},
  title        = {The probabilistic model checker {S}torm: symbolic methods for probabilistic model checking},
  school       = {{RWTH} Aachen University, Germany},
  year         = {2018},
  url          = {http://publications.rwth-aachen.de/record/752011},
  urn          = {urn:nbn:de:101:1-2019060410510882734859}
}

@inproceedings{BDH+17,
  author       = {Carlos E. Budde and
                  Christian Dehnert and
                  Ernst Moritz Hahn and
                  Arnd Hartmanns and
                  Sebastian Junges and
                  Andrea Turrini},
  editor       = {Axel Legay and
                  Tiziana Margaria},
  title        = {{JANI:} Quantitative Model and Tool Interaction},
  booktitle    = {23rd International Conference on Tools and Algorithms for the Construction and Analysis of Systems ({TACAS} 2017)},
  series       = {Lecture Notes in Computer Science},
  volume       = {10206},
  pages        = {151--168},
  year         = {2017},
  doi          = {10.1007/978-3-662-54580-5\_9}
}

@inproceedings{BHHK15,
  author       = {Yuliya Butkova and
                  Hassan Hatefi and
                  Holger Hermanns and
                  Jan Krc{\'{a}}l},
  editor       = {Bernd Finkbeiner and
                  Geguang Pu and
                  Lijun Zhang},
  title        = {Optimal Continuous Time {M}arkov Decisions},
  booktitle    = {13th International Symposium on Automated Technology for Verification and Analysis ({ATVA} 2015)},
  series       = {Lecture Notes in Computer Science},
  volume       = {9364},
  pages        = {166--182},
  publisher    = {Springer},
  year         = {2015},
  doi          = {10.1007/978-3-319-24953-7\_12}
}

@inproceedings{DJLMT15,
  author       = {Alexandre David and
                  Peter Gj{\o}l Jensen and
                  Kim Guldstrand Larsen and
                  Marius Mikucionis and
                  Jakob Haahr Taankvist},
  editor       = {Christel Baier and
                  Cesare Tinelli},
  title        = {Uppaal {S}tratego},
  booktitle    = {21st International Conference on Tools and Algorithms for the Construction and Analysis of Systems ({TACAS} 2015)},
  series       = {Lecture Notes in Computer Science},
  volume       = {9035},
  pages        = {206--211},
  publisher    = {Springer},
  year         = {2015},
  doi          = {10.1007/978-3-662-46681-0\_16}
}

@inproceedings{HH15,
  author       = {Arnd Hartmanns and
                  Holger Hermanns},
  editor       = {Bernd Finkbeiner and
                  Geguang Pu and
                  Lijun Zhang},
  title        = {Explicit Model Checking of Very Large {MDP} Using Partitioning and Secondary Storage},
  booktitle    = {13th International Symposium on Automated Technology for Verification and Analysis ({ATVA} 2015)},
  series       = {Lecture Notes in Computer Science},
  volume       = {9364},
  pages        = {131--147},
  publisher    = {Springer},
  year         = {2015},
  doi          = {10.1007/978-3-319-24953-7\_10}
}

@inproceedings{HH14,
  author       = {Arnd Hartmanns and
                  Holger Hermanns},
  editor       = {Erika {\'{A}}brah{\'{a}}m and
                  Klaus Havelund},
  title        = {The {M}odest {T}oolset: An Integrated Environment for Quantitative Modelling and Verification},
  booktitle    = {20th International Conference on Tools and Algorithms for the Construction and Analysis of Systems ({TACAS} 2014)},
  series       = {{LNCS}},
  volume       = {8413},
  pages        = {593--598},
  publisher    = {Springer},
  year         = {2014},
  doi          = {10.1007/978-3-642-54862-8\_51}
}

@article{GLMS13,
  author       = {Hubert Garavel and
                  Fr{\'{e}}d{\'{e}}ric Lang and
                  Radu Mateescu and
                  Wendelin Serwe},
  title        = {{CADP} 2011: a toolbox for the construction and analysis of distributed processes},
  journal      = {Int. J. Softw. Tools Technol. Transf.},
  volume       = {15},
  number       = {2},
  pages        = {89--107},
  year         = {2013},
  doi          = {10.1007/S10009-012-0244-Z}
}

@article{HHHK13,
  author       = {Ernst Moritz Hahn and
                  Arnd Hartmanns and
                  Holger Hermanns and
                  Joost-Pieter Katoen},
  title        = {A compositional modelling and analysis framework for stochastic hybrid systems},
  journal      = {Formal Methods Syst. Des.},
  volume       = {43},
  number       = {2},
  pages        = {191--232},
  year         = {2013},
  doi          = {10.1007/S10703-012-0167-Z}
}

@inproceedings{GHKN12,
  author       = {Dennis Guck and
                  Tingting Han and
                  Joost-Pieter Katoen and
                  Martin R. Neuh{\"{a}}u{\ss}er},
  editor       = {Alwyn Goodloe and
                  Suzette Person},
  title        = {Quantitative Timed Analysis of Interactive {M}arkov Chains},
  booktitle    = {4th International {NASA} Formal Methods Symposium ({NFM} 2012)},
  series       = {Lecture Notes in Computer Science},
  volume       = {7226},
  pages        = {8--23},
  publisher    = {Springer},
  year         = {2012},
  doi          = {10.1007/978-3-642-28891-3\_4}
}

@inproceedings{KNP11,
  author       = {Marta Z. Kwiatkowska and
                  Gethin Norman and
                  David Parker},
  editor       = {Ganesh Gopalakrishnan and
                  Shaz Qadeer},
  title        = {{PRISM} 4.0: Verification of Probabilistic Real-Time Systems},
  booktitle    = {23rd International Conference on Computer Aided Verification ({CAV} 2011)},
  series       = {Lecture Notes in Computer Science},
  volume       = {6806},
  pages        = {585--591},
  publisher    = {Springer},
  year         = {2011},
  doi          = {10.1007/978-3-642-22110-1\_47}
}

@inproceedings{EHZ10,
  author       = {Christian Eisentraut and
                  Holger Hermanns and
                  Lijun Zhang},
  title        = {On Probabilistic Automata in Continuous Time},
  booktitle    = {25th Annual {IEEE} Symposium on Logic in Computer Science ({LICS} 2010)},
  pages        = {342--351},
  publisher    = {{IEEE} Computer Society},
  year         = {2010},
  doi          = {10.1109/LICS.2010.41}
}

@book{BK08,
  author       = {Christel Baier and
                  Joost-Pieter Katoen},
  title        = {Principles of model checking},
  publisher    = {{MIT} Press},
  year         = {2008}
}

@article{BDHK06,
  author       = {Henrik C. Bohnenkamp and
                  Pedro R. D'Argenio and
                  Holger Hermanns and
                  Joost-Pieter Katoen},
  title        = {{MoDeST}: A Compositional Modeling Formalism for Hard and Softly Timed Systems},
  journal      = {{IEEE} Trans. Software Eng.},
  volume       = {32},
  number       = {10},
  pages        = {812--830},
  year         = {2006},
  doi          = {10.1109/TSE.2006.104}
}

@book{Saa03,
  author       = {Yousef Saad},
  title        = {Iterative Methods for Sparse Linear Systems},
  publisher    = {{SIAM}},
  year         = {2003}
}

@article{KNSS02,
  author       = {Marta Z. Kwiatkowska and
                  Gethin Norman and
                  Roberto Segala and
                  Jeremy Sproston},
  title        = {Automatic verification of real-time systems with discrete probability distributions},
  journal      = {Theor. Comput. Sci.},
  volume       = {282},
  number       = {1},
  pages        = {101--150},
  year         = {2002},
  doi          = {10.1016/S0304-3975(01)00046-9}
}

@article{GLD00,
  author       = {Robert Givan and
                  Sonia M. Leach and
                  Thomas L. Dean},
  title        = {Bounded-parameter {M}arkov decision processes},
  journal      = {Artif. Intell.},
  volume       = {122},
  number       = {1-2},
  pages        = {71--109},
  year         = {2000},
  doi          = {10.1016/S0004-3702(00)00047-3}
}

@article{KLC98,
  author       = {Leslie P. Kaelbling and
                  Michael L. Littman and
                  Anthony R. Cassandra},
  title        = {Planning and Acting in Partially Observable Stochastic Domains},
  journal      = {Artif. Intell.},
  volume       = {101},
  number       = {1-2},
  pages        = {99--134},
  year         = {1998},
  doi          = {10.1016/S0004-3702(98)00023-X}
}

@book{Put94,
  author       = {Martin L. Puterman},
  title        = {{M}arkov Decision Processes: Discrete Stochastic Dynamic Programming},
  series       = {Wiley Series in Probability and Statistics},
  publisher    = {Wiley},
  year         = {1994},
  doi          = {10.1002/9780470316887},
  isbn         = {978-0-47161977-2}
}

@article{Con92,
  author       = {Anne Condon},
  title        = {The Complexity of Stochastic Games},
  journal      = {Inf. Comput.},
  volume       = {96},
  number       = {2},
  pages        = {203--224},
  year         = {1992},
  doi          = {10.1016/0890-5401(92)90048-K}
}

@article{Sha53,
  author       = {Lloyd S. Shapley},
  title        = {Stochastic games},
  journal      = {Proceedings of the National Academy of Sciences},
  volume       = {39},
  number       = {10},
  pages        = {1095--1100},
  year         = {1953},
  doi          = {10.1073/pnas.39.10.1095}
}
